\documentclass[reprint, superscriptaddress, amsmath, amssymb, aps, pra]{revtex4-2}
\usepackage{graphicx, placeins, afterpage}
\usepackage{dcolumn}%
\usepackage{xcolor}
\usepackage[absolute]{textpos}
\definecolor{darkblue}{rgb}{0,0.3,0.7}
\usepackage{hyperref, comment}
\hypersetup{
colorlinks=true,
linkcolor=darkblue,
filecolor=blue,
citecolor=darkblue,  
urlcolor=darkblue,
}

\definecolor{ASUblue}{RGB}{0,163,224}

\begin{document}

\preprint{APS/123-QED}

\title{On-chip pulse shaping of entangled photons}
\author{Kaiyi Wu}
\email{wu1871@purdue.edu}
\thanks{equal contribution.}
\affiliation{Elmore Family School of Electrical and Computer Engineering and Purdue Quantum Science and Engineering Institute, Purdue University, West Lafayette, Indiana 47907, USA}

\author{Lucas M. Cohen}
\email{cohen26@purdue.edu}
\thanks{equal contribution.}
\affiliation{Elmore Family School of Electrical and Computer Engineering and Purdue Quantum Science and Engineering Institute, Purdue University, West Lafayette, Indiana 47907, USA}

\author{Karthik V. Myilswamy}
\affiliation{Elmore Family School of Electrical and Computer Engineering and Purdue Quantum Science and Engineering Institute, Purdue University, West Lafayette, Indiana 47907, USA}

\author{Navin B. Lingaraju}
\affiliation{The Johns Hopkins University Applied Physics Laboratory, Laurel, Maryland 20723, USA}

\author{Hsuan-Hao Lu}
\affiliation{Quantum Information Science Section, Computational Sciences and Engineering Division, Oak Ridge National Laboratory, Oak Ridge, Tennessee 37831, USA}

\author{Joseph~M. Lukens}
\affiliation{Quantum Information Science Section, Computational Sciences and Engineering Division, Oak Ridge National Laboratory, Oak Ridge, Tennessee 37831, USA}
\affiliation{Research Technology Office and Quantum Collaborative, Arizona State University, Tempe, Arizona 85287, USA}

\author{Andrew~M. Weiner}
 \affiliation{Elmore Family School of Electrical and Computer Engineering and Purdue Quantum Science and Engineering Institute, Purdue University, West Lafayette, Indiana 47907, USA}
\thanks{Legacy.}

\date{\today}

\begin{abstract}
We demonstrate spectral shaping of entangled photons with a six-channel microring-resonator-based silicon photonic pulse shaper. Through precise calibration of thermal phase shifters in a microresonator-based pulse shaper, we demonstrate line-by-line phase control on a 3~GHz grid for two frequency-bin-entangled
qudits, corresponding to Hilbert spaces of up to $6\times 6$ ($3\times 3$) dimensions for shared (independent) signal-idler filters. The pulse shaper's fine spectral resolution enables control of nanosecond-scale temporal features, which are observed by direct coincidence detection of biphoton correlation functions that show excellent agreement with theory. This work marks, to our knowledge, the first demonstration of biphoton pulse shaping using an integrated spectral shaper and holds significant promise for applications in quantum information processing.
\end{abstract}

\maketitle

\begin{figure*}[hbt!]
\centerline{\includegraphics[width=0.95\textwidth]{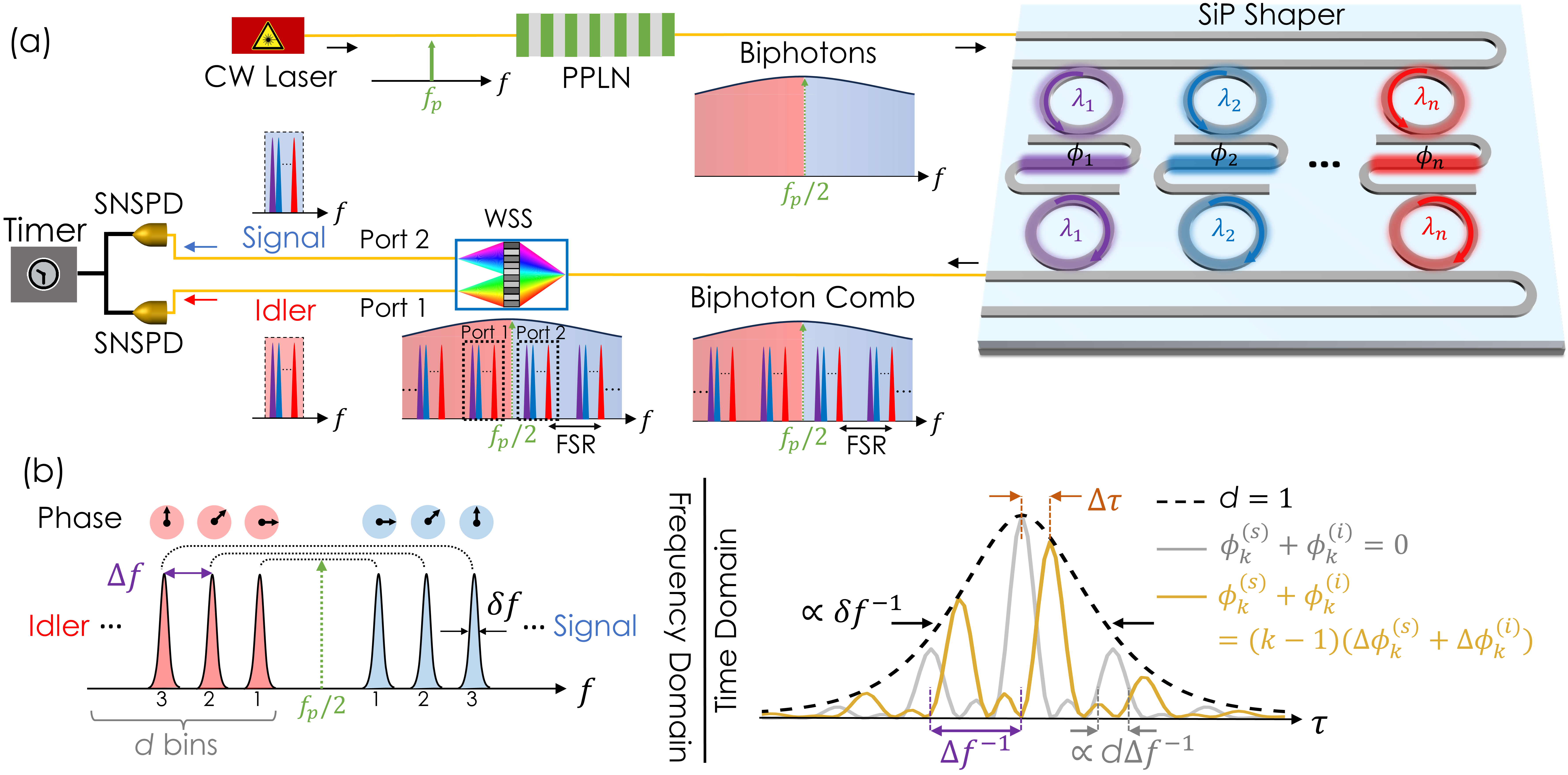}}
\caption{
(a) Diagram of the experimental setup. CW: continuous-wave. PPLN: periodically poled lithium niobate. SiP: silicon photonic. FSR: free spectral range. SNSPD: superconducting nanowire single-photon detector. WSS: wavelength-selective switch.
(b) Conceptual illustration of high-dimensional frequency-bin states, depicting $d$ coherent superpositions of frequency-bin pairs and their corresponding temporal correlation functions. In this example, linear spectral phases are applied to both signal and idler bins, resulting in a temporal offset (gold) compared to the case with constant phase (gray). See text for details. 
}
\label{fig1}
\end{figure*}

\section{Introduction}
Entanglement serves as a pivotal resource in quantum information processing (QIP)~\cite{Nielsen2000, monroe2002quantum,flamini2018photonic}, quantum communication~\cite{Gisin2002, Lo2014, LuCao2022,
luo2023recent}, and quantum networking systems~\cite{munro2010quantum,Wehner2018}. Entangled photons, due to their compatibility with the classical optical telecommunications infrastructure and resilience against decoherence, are omnipresent carriers of quantum information. In recent years, photonic frequency-bin entanglement has garnered significant attention because of its inherent high dimensionality, which is valuable for the scaling of quantum systems~\cite{Erhard2020}. These high-dimensional frequency-bin entangled photon pairs, known as biphoton frequency combs (BFCs), can be produced through spontaneous nonlinear optical processes~\cite{kues2019quantum, lu2022bayesian,lu2023frequency}, 
such as comb-like spectral filtering of broadband
biphoton spectra from parametric downconversion and direct pumping of parametric oscillators like integrated resonators well below threshold.

To fully realize QIP with BFCs, one must be able to both manipulate and measure the quantum state of the BFC in a line-by-line fashion.
A critical piece of this objective is the Fourier-transform pulse shaper, which typically leverages a liquid crystal-based spatial light modulator sandwiched between two dispersive elements (e.g., diffraction gratings or prisms) to synthesize arbitrary spectral filters~\cite{Weiner2000, Weiner2011}. 
Today a permanent fixture of classical optical applications---ranging from spectroscopy~\cite{Silberberg2009} to RF arbitrary waveform generation~\cite{Torres2014}, coherent control of chemical reactions~\cite{Assion1998} to lightwave communications~\cite{Roelens2008}---pulse shapers have likewise made their impact felt in the quantum domain. Initial experiments shaping the correlation function of entangled photons with homebuilt free-space devices~\cite{Peer2005, Zaeh2008, bernhard2013shaping} have been followed by commercial fiber-pigtailed pulse shapers supporting spectral control with $\sim$10~GHz resolution~\cite{Lukens2013b, Lukens2013c, lukens2014orthogonal}, as well as spatial control when configured as multioutput wavelength-selective switches (WSSs)~\cite{Lingaraju2021, Appas2021}. Combined with electro-optic phase modulators via the quantum frequency processor paradigm, such pulse shapers can even enable universal QIP in frequency-bin encoding with BFCs as resource states~\cite{Lukens2017, lu2023frequency}.  Additionally, optical pulse shaping has facilitated temporal-mode QIP~\cite{Brecht2015, Ansari2018}, playing a critical role in the synthesis of quantum pulse gates through the spectral shaping of classical pump fields~\cite{Brecht2014, Serino2023}. 

Yet despite pulse shaping's expanding importance to quantum photonics, existing bulk solutions are reaching their limits in terms of size ($\sim$1~m$^2$) 
and spectral resolution ($\gtrsim$10~GHz), thereby constraining prospects for the continued scaling of BFC processing systems to new regimes of dimensionality and circuit depth. Inspired by a handful of prior examples in silicon photonics~\cite{Agarwal2006, Khan2010, Wang2015b}, alternative pulse shaping designs based on microring resonators have been proposed to surmount this impasse. In this concept, microring filters coupled to an input bus waveguide download individual spectral lines, apply phase shifts, and then upload them back to the output waveguide. 
Nevertheless, pulse shaping of nonclassical light with such a design has proven elusive, due in large part to the technical challenges associated with calibrating and controlling microring filter banks.

In this paper, we overcome these challenges and demonstrate on-chip spectral shaping of entangled photons. Our silicon-microring-based pulse shaper features six spectral channels, each of which is simultaneously calibrated and stabilized via a recently developed multiheterodyne, dual-comb technique~\cite{Cohen2024nc}. Carving the spectrum of broadband input biphotons, 
the pulse shaper realizes line-by-line phase shaping down to resolutions as fine as $\sim$3~GHz---a record for biphoton pulse shaping that facilitates observation of shaped biphoton wavepackets directly in the time domain. 
Two regimes of pulse shaper operation are highlighted: spacing the signal and idler channels at a multiple of the microring free spectral range (FSR) permits shaping of two entangled qudits up to six dimensions in each photon (i.e., $d=6$), but with correlations in the signal and idler phases, whereas shifting the signal and idler channels to separate FSR grids enables fully arbitrary phase control, but up to two qutrits only ($d=3$). In all cases, results show excellent agreement with theory, confirming the feasibility of integrated pulse shapers for quantum state control.
By incorporating additional microrings with higher quality factors, our basic design can be extended to more channels and finer resolution, as well as combined with other CMOS quantum photonic devices for all-in-one processing circuits on chip---circuits that open new opportunities not only in frequency-bin QIP, but also in any context where precise spectro-temporal mode matching of interacting photons is required.

\section{Theory and Experimental setup}
\label{sec:theoryAndSetup}
The experimental setup is depicted in Fig.~\ref{fig1}(a).
We obtain broadband time-energy entangled photons for testing via type-0 spontaneous parametric downconversion (SPDC) in a periodically poled lithium niobate (PPLN) waveguide driven by a continuous-wave (CW) laser operating at frequency $f_p\approx 386.8$~THz (775~nm). We couple the biphotons centered at 193.4~THz (1550~nm) onto our silicon photonic (SiP) shaper chip, whose design and programming methods are described in detail in our previous work shaping classical optical frequency combs~\cite{Cohen2024nc}. 
The SiP shaper contains a common input waveguide that first directs the broadband biphotons to a microresonator filter bank that downloads spectral slices, applies a phase shift to each with a resistive heater, and then uploads the spectrum to a common output waveguide through a second microresonator filter bank. 

The spectral shaper contains six channels with a full-width at half-maximum (FWHM) bandwidth $\delta f = \delta\omega/2\pi \approx 900$~MHz (after both microrings), FSR $f_\mathrm{FSR} = \omega_\mathrm{FSR}/2\pi \approx 115$~GHz, and an insertion loss of $\sim$6~dB (excluding fiber-to-chip coupling loss, which we measure to be $\sim$3.5~dB/facet). In our experiment, the pulse shaper both creates the BFC and tunes its phase, but could just as easily shape an independently generated BFC of appropriate spacing.
At the output of the SiP shaper, a commercial WSS (WaveShaper 4000S; Finisar) with 10~GHz resolution spectrally isolates and spatially separates the shaped signal and idler bins, symmetrically carved about the center of the biphoton spectrum. 
Each photon is then routed to a superconducting nanowire single-photon detector (SNSPD; Quantum Opus Opus One) connected to a timing module (Time Tagger Ultra; Swabian) for time-resolved coincidence detection.

To model the shaped photonic state, we assume each microresonator filter is tuned to a grid spaced by $\Delta\omega$, making the center frequencies $\omega_k^{(s)} = \frac{\omega_p}{2} + (k+B)\Delta\omega$ for the $k^\mathrm{th}$ signal bin and $\omega_k^{(i)} = \frac{\omega_p}{2} - (k+B)\Delta\omega$ for the $k^\mathrm{th}$ idler bin, 
where $B$ is a positive number (not necessarily integer) describing how far the resonances of interest lie from the center of the biphoton spectrum. Under the assumption of nonoverlapping peaks, the complex transfer functions for the signal and idler are

\begin{equation}
\begin{split}
    H_s(\omega) & = \sum_{k=1}^{d} \frac{e^{i\phi_k^{(s)}}}{\left[\frac{\gamma}{2} + i(\omega-\omega_{k}^{(s)})\right]^2}\\  
    H_i(\omega) & = \sum_{k=1}^{d} \frac{e^{i\phi_k^{(i)}}}{\left[\frac{\gamma}{2} + i(\omega-\omega_{k}^{(i)})\right]^2},
\end{split}
\label{eq_filter}
\end{equation}
where $\phi_k^{(s)}$ and $\phi_k^{(i)}$ represent the phase implemented using the inline phase shifter for the $k^\mathrm{th}$ signal and idler bin, respectively. Each term follows a Lorentzian lineshape 
centered at the appropriate target signal frequency $\omega_{k}^{(s)}$ or idler frequency $\omega_{k}^{(i)}$, and is squared to account for both the download and upload rings. Hence the channel FWHM $\delta\omega$ is slightly smaller than the single-ring FWHM $\gamma$---namely, $\delta\omega = 0.644\gamma$.
The dimensionality $d$ of the BFC corresponds to the number of physical channels acting on the signal and idler photons.

In Eq.~(\ref{eq_filter}), we have assumed identical resonance frequencies for the download and upload rings of each channel.
The signal (idler) bin closest to the center of the biphoton spectrum is $k=1$, with the bins at higher (lower) frequencies numbered sequentially as $\{2,\hdots,d\}$. 
This ordering ensures that the signal and idler bins which are correlated due to the energy conservation of the SPDC process are assigned the same number, i.e., $\omega_{k}^{(s)}+\omega_{k}^{(i)} = \omega_p \; \forall \; k$. An illustration of the BFC spectrum carved by the shaper in this way is shown in the left half of Fig.~\ref{fig1}(b). 

The temporal wavepacket of the BFC can be expressed as~\cite{lukens2014generation}:
\begin{equation}
    \psi(\tau) = \int_0^\infty d\Omega \, \Phi (\Omega) H_s\left(\frac{\omega_p}{2} + \Omega\right) H_i\left(\frac{\omega_p}{2} - \Omega\right) e^{-i\Omega \tau},
    \label{eq_corr}
\end{equation}
where $\tau$ represents the time delay between signal and idler photons, $H_s(\omega)$ and $H_i(\omega)$ are the complex spectral transfer functions 
defined by Eq.~(\ref{eq_filter}), and $\Phi(\Omega)$ describes the spectral amplitude of the initial broadband biphoton generated through SPDC, governed by the phase-matching conditions of the nonlinear waveguide. Over the bandwidths of our pulse shaper, we can safely assume $\Phi(\Omega) = 1$ (i.e., constant with uniform phase). 

Under these conditions, we obtain (neglecting unobservable unimodular factors)
\begin{equation}
\label{eq:wavepacket}
\psi(\tau) = \sum_{k=1}^d e^{i(\phi_k^{(s)}+\phi_k^{(i)}-k\Delta\omega\tau)} \int_{-\infty}^\infty d\Omega \frac{e^{-i\Omega\tau}}{\left( \frac{\gamma^2}{4} + \Omega^2 \right)^2}, 
\end{equation}
where we have exploited the nonoverlapping resonance approximation to eliminate all but the energy-matched peaks from the expression.
Significantly, the biphoton's frequency correlations cause the phase of the Lorentzian lineshapes to cancel completely, leaving the only nontrivial spectral phases as those applied explicitly by the pulse shaper ($\phi_k^{(s)}$ and $\phi_k^{(i)}$). Of course, such cancellation does not occur for generic quantum states; the impact of the full phase response has been incorporated in prior analyses of integrated pulse shapers and found consistent with high-fidelity frequency-bin operations, for example~\cite{Nussbaum2022}. Nevertheless, its absence here simplifies both modeling and comprehension of our experimental results.

The correlation function defined as $G^{(2)}(\tau) = |\psi(\tau)|^2$ is proportional to the coincidence rate at delay $\tau$: i.e., the probability of detecting the signal photon delayed by $\tau$ with respect to the idler over some infinitesimal window $d\tau$, for detectors fast enough to resolve all temporal features.
An example correlations function is portrayed in the right half of Fig.~\ref{fig1}(b) which, following Eq.~(\ref{eq:wavepacket}), comprises interferometric terms (in the sum) oscillating at the interchannel spacing that are multiplied by an envelope (the integral) whose duration is determined by the inverse linewidth of a single channel.
For net zero biphoton phase, i.e. $\phi_k^{(s)} + \phi_k^{(i)} = 0$, the wavepacket shows 
a peak centered at $\tau=0$ (depicted in gray). The duration of a single fringe is inversely proportional to the total bandwidth of the signal (or idler) half of BFC, or $d\Delta f$. Analogous to the effect in classical optical pulse shaping, we can then manipulate the biphoton wavepacket through the programmable line-by-line phase control of our SiP shaper. For example, applying the linear phases $\phi_k^{(s)}=(k-1)\Delta\phi^{(s)}$ and $\phi_k^{(i)}=(k-1)\Delta\phi^{(i)}$ 
causes the fringes under the envelope to shift by an amount $\Delta\tau = (\Delta\phi_k^{(s)}+\Delta\phi_k^{(i)})/\Delta\omega$, as in the gold case in Fig.~\ref{fig1}(b).

\begin{figure*}[ht!]
\centerline{\includegraphics[width=\textwidth]{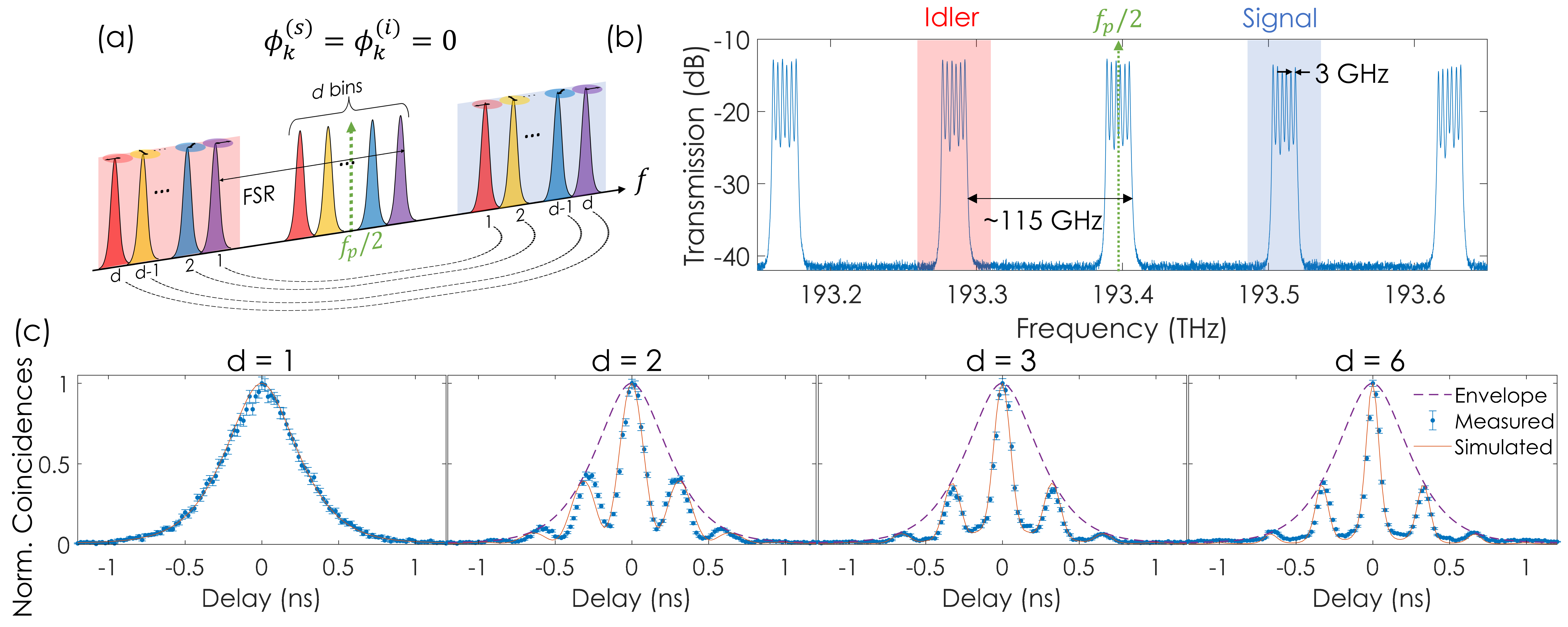}}
\caption{
Shared signal-idler filters. (a) Conceptual diagram and phase settings. See Sec.~\ref{ExpResults}-A for details. (b) Normalized transmission spectrum of the SiP shaper chip with six channels programmed. The blue and red shaded regions indicate the passbands programmed on the WSS. (c) Normalized signal-idler temporal correlation functions for various BFC dimensions $d$. The blue scatter plot is from measurement, the orange curve is from simulation, and the purple curve is the envelope from the simulated temporal correlation at $d=1$. All error bars denote standard deviations assuming Poissonian statistics. 
}
\label{fig2}
\end{figure*}

After spatially separating the shaped signal and idler photons we perform time-resolved coincidence detection; all experimental results are obtained with 20~ps histogram bins. 
According to theory, the measured coincidence rate $C(\tau)$ for any configuration is given by the convolution 
\begin{equation}
\label{eq:convolve}
C(\tau) \propto \int_{-\infty}^\infty d\tau'\, G^{(2)}(\tau')h(\tau-\tau'),
\end{equation}
where $h(\tau)$ denotes the system impulse response (detectors and timetagger). We measure $h(\tau)$ by bypassing the SiP shaper and transmitting $50$~GHz-wide slots from the WSS, chosen sufficiently broad to ensure $C(\tau) \propto h(\tau)$; $h(\tau)$ is well approximated by a Gaussian with an 80~ps FWHM (see Appendix~\ref{app:A} for details). All simulation curves below are calculated via Eq.~(\ref{eq:convolve}) using this function.

Because the biphoton flux from our PPLN waveguide setup fluctuates significantly due to unstable coupling, we normalize all experimental correlation functions for ease of comparison. Specifically, histograms of the same qudit dimension $d$ are scaled to conserve the total area under each curve---justified since the tested configurations differ only in spectral phase~\cite{myilswamy2023time}. The peak across a given a family of scaled curves is then assigned a value of one for plotting in dimensionless units.



\section{Experimental Results}
\label{ExpResults}
\subsection{Shared Signal-Idler Filters}
\label{sec:shared}
We first demonstrate shaping for several dimensions $d$ to highlight the impact of bin number on the  time-resolved features of the biphoton correlation function. We program the SiP shaper to pick out $d\in\{1,2,3,6\}$ bins at $\Delta f = \Delta\omega/2\pi =3$~GHz bin spacing. Zero phase is programmed across the bins for each dimension ($d=1$~and~$d=2$ have only global or linear phase making them trivial cases). Details on phase programming can be found in Appendix~\ref{app:B}. Our method involving linear phase subtraction by setting two of the channels as a linear phase reference may result in a global linear phase across all channels (beyond the group delay through the device itself), which has no impact on the measured biphoton wavepacket. As depicted in Fig.~\ref{fig2}(a), the center of the biphoton spectrum ($f_p/2$) is aligned to the midpoint of the bins in one FSR of the shaper, and signal (idler) bins are carved at one FSR above (below). Figure~\ref{fig2}(b) shows the measured transmission spectrum of the SiP shaper for the $d=6$ case, where the shaded regions denote the WSS passbands used to spatially separate the signal and idler bins for measurement.  For the cases where $d<6$, the rings in the unused channels are detuned to fall outside of the WSS passband.

In this first configuration, the signal and idler bins of interest are separated by an integer multiple of the microring FSR. Consequently, the $k^\mathrm{th}$ bin of the idler traverses the same physical microring filter as the $(d-k+1)^\mathrm{th}$ signal and hence experiences the same phase: $\phi_k^{(i)} = \phi_{d-k+1}^{(s)}$.
If there is a linear phase across the pulse shaper channels such that $\phi_k^{(s)} = (k-1)\Delta\phi$, the idler phases follow as $\phi_k^{(i)} = (d-k)\Delta\phi$ and $\phi_k^{(s)}+\phi_k^{(i)}=(d-1)\Delta\phi$. In other words, any linear phase on the pulse shaper channels reduces to a constant in Eq.~(\ref{eq:wavepacket}), implying a correlation function centered at $\tau=0$.

\begin{figure*}[htb!]
\centerline{\includegraphics[width=\textwidth]{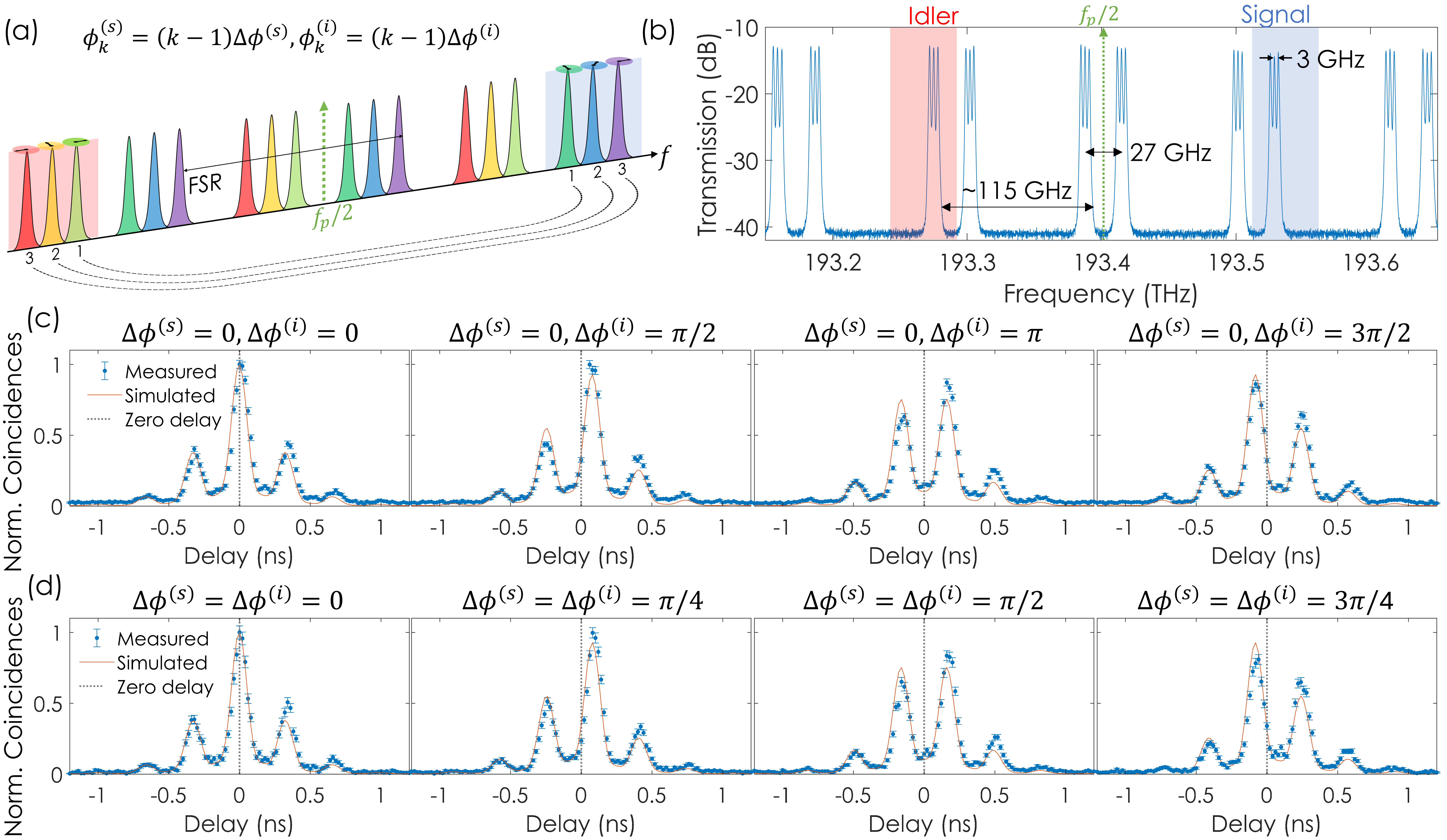}}
\caption{Distinct signal-idler filters. (a) Conceptual diagram and phase settings. See Sec.~\ref{ExpResults}-B for details. (b) Normalized transmission spectrum of the SiP shaper chip programmed to manipulate two subcombs of three bins. The blue and red shaded regions indicate the passbands programmed on the WSS. (c) Normalized signal-idler temporal correlation functions when $\Delta \phi^{(s)} = 0,\,\Delta \phi^{(i)} \neq 0$. (d)~Normalized signal-idler temporal correlation functions when $\Delta \phi^{(s)}=\Delta \phi^{(i)} \neq 0$. Error bars assume Poissonian statistics.
}
\label{fig3}
\end{figure*}


Figure~\ref{fig2}(c) shows the measured (blue scatter plot) and simulated (orange line) biphoton correlation function, as well as the envelope obtained from the $d=1$ case (purple line). For all simulations, we take $\gamma/2\pi = 1.3$~GHz (the value returned from curve fitting the $d=1$ case). As $d$ increases, pulse-like fringe features caused by interference between bins appear within the envelope of the $d=1$ case. As expected, these fringes oscillate at a period of $\sim$320~ps, equal (to within histogram resolution) to the inverse of the 3~GHz bin spacing. As $d$ increases, the width of the fringes narrows and the extinction between them becomes stronger. Although detector jitter reduces the contrast slightly from 100\% between peaks, the ultrafine 3~GHz pulse shaper resolution nevertheless leads to temporal features clearly resolvable with our photon detectors, showing excellent agreement with theoretical predictions from Eq.~(\ref{eq:convolve}).

\subsection{Distinct Signal-Idler Filters}
\label{sec:distinct}
Tuning the microring resonances such that the signal and idler bins of interest pass through the same physical filters maximizes the total number of accessible signal or idler bins (six in our case), but limits the level of controllability, due to the restriction $\phi_k^{(i)} = \phi_{d-k+1}^{(s)}$. Yet independent control of signal-idler phase is crucial for various applications, such as implementing distinct quantum gates for photons traveling in the same fiber~\cite{Lu2018b} or realizing spectral coding methods on time-frequency entangled photons~\cite{lukens2014orthogonal}. 

To apply fully independent spectral phase on the bins for each photon, we next program the SiP shaper to transmit two subcombs of $d=3$ bins at the same $3$~GHz spacing, with a subcomb center-to-center spacing of $27$~GHz
---a multiple of 3~GHz to facilitate phase calibration with our dual-comb heterodyne technique (cf. Appendix~\ref{app:B}). We then retune the pump laser to align the center of the biphoton spectrum to the midpoint of these two subcombs. As shown schematically in Fig.~\ref{fig3}(a), the signal and idler bins now correspond to different subcombs; for the signal photon, the higher-frequency subcomb in its respective FSR is used, while for the idler photon, the lower-frequency subcomb is utilized. This scheme allows us to apply arbitrary phases to both the signal and idler portions of a $d=3$ BFC, as each signal and idler bin is travels through a unique physical channel.
The measured transmission spectrum of the SiP shaper in this configuration is shown in Fig.~\ref{fig3}(b), with shading highlighting the 50~GHz-wide WSS passbands used for demultiplexing.
 
\begin{figure*}[htb!]
\centerline{\includegraphics[width=\textwidth]{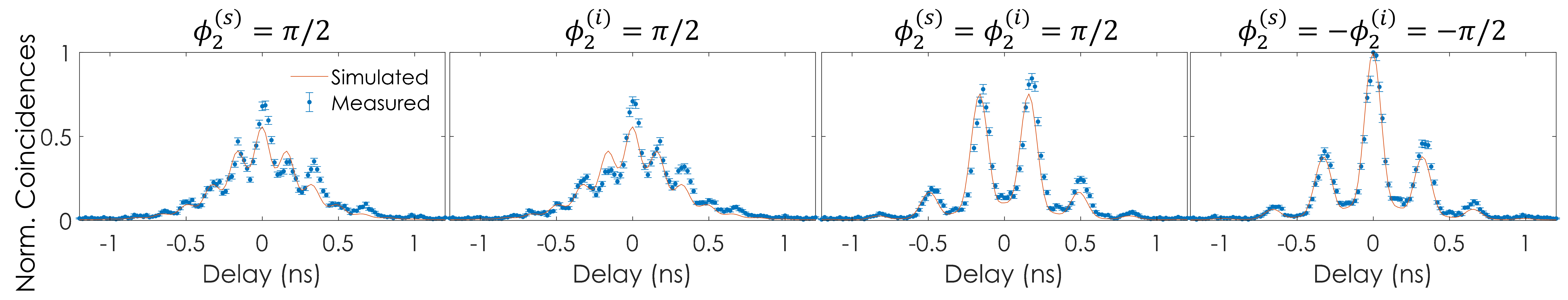}}
\caption{Normalized temporal correlations measured for distinct signal-idler filters with nonlinear phases. In each case, all phases are set to zero [$\phi_k^{(s,i)}=0$] except for those listed above the respective plot. Error bars assume Poissonian statistics.
}
\label{fig4}
\end{figure*}

Consider the linear phases $\phi_k^{(s)}=(k-1)\Delta\phi^{(s)}$ and $\phi_k^{(i)}=(k-1)\Delta\phi^{(i)}$ applied across the signal and idler bins, respectively. As noted in Sec.~\ref{sec:theoryAndSetup}, these values
shift the fringes under the envelope by $\Delta\tau = (\Delta\phi^{(s)}+\Delta\phi^{(i)})/\Delta\omega$. To study this effect experimentally, we consider linear phase applied to the idler photon ($\Delta\phi^{(i)} \neq 0$) with the signal experiencing either zero phase ($\Delta \phi^{(s)} = 0$) or linear phase of the same slope ($\Delta \phi^{(s)} = \Delta\phi^{(i)}$). More details on phase programming of this filter spectrum can be found in the Appendix. 

Figure~\ref{fig3}(c) shows the measured and simulated temporal correlation functions for various linear phase increments across the idler  $\Delta\phi^{(i)} \in\{0,\pi/2, \pi, 3\pi/2\}$ with the signal fixed at $\Delta\phi^{(s)} = 0$; Fig.~\ref{fig3}(d) covers the case with the same phase increments on both photons, where $\Delta\phi^{(s)}=\Delta\phi^{(i)} \in \{0, \pi/4, \pi/2, 3\pi/4\}$. Note that the phase increments in the second case are halved from the first to produces the same sums $\phi_k^{(s)}+\phi_k^{(i)}$, ideally leading to the same correlation functions despite the distinct physical configurations.

As expected, the fringes shift under the envelope by an amount corresponding to the sum $\Delta\phi^{(s)}+\Delta\phi^{(i)}$ rather than individual slopes, displaying good agreement with simulated predictions for all cases. 
For the cases with $\Delta \phi^{(s)} + \Delta\phi^{(i)} = \pi$, we see slight asymmetry in the measured correlation functions, which we attribute to coherent crosstalk from the finite resolution of our SiP shaper. This crosstalk plays an effect in each measurement but is more prominent for these phase configurations.

\subsection{Nonlinear Phases}
\label{sec:nonlinear}
Beyond special cases of linear phase which lead to fringe delays, the line-by-line capabilities of the SiP pulse shaper enable more general manipulation of  the correlation function as well. To demonstrate this, we proceed with our carved $3$-bin subcombs but allow only the phase of the central bin of each subcomb ($\phi_2^{(s,i)}$) to vary. The first and second plots in Fig.~\ref{fig4} show the correlation functions when a $\pi/2$ phase is applied across the middle signal and idler bin, respectively, while all the other bins are fixed at zero relative phase. The correlation function shows a doubled repetition rate for these cases as expected from theoretical simulations, though the doubled repetition rate puts the oscillation period ($\sim$170~ps) closer to the system response ($\sim$80~ps), thus reducing the contrast.

Next, we apply a $\pi/2$ phase to the middle bin of \emph{both} signal and idler subcombs, with the results shown in the third plot of Fig.~\ref{fig4}. The ideal biphoton phase in this case is identical to the third column in Fig.~\ref{fig3}(d) and, indeed, we see a similar shape to the correlation function with a clear null at $\tau=0$. Finally, we show that the phase applied to one photon can be compensated for by applying an appropriate phase to the other photon by setting the middle signal and idler bin to $- \pi/2$ and $\pi/2$, respectively (last plot of Fig.~\ref{fig4}). As expected, we see very good agreement between the measured and simulated results, as well as with the same state explored in other configurations shown in Fig.~\ref{fig2}(c) (third column) and Fig.~\ref{fig3}(c,d) (first column).

\section{Discussion and Conclusion}
In this paper, we showed the use of a high-resolution spectral shaper for manipulating narrowband frequency-bin-entangled photons. Using our chip, we carved tightly spaced BFCs from a broadband SPDC spectrum and applied controllable phases to different bins. The narrow bin widths realized by our SiP shaper produce biphotons with broad temporal wavepackets, thereby enabling simple time-resolved measurement with commercial single-photon detectors. We first observed the effect of increasing the BFC dimension (number of bins) on the temporal correlation function. Next, we showed the ability to control the spectral phase of bins which carve either or both the signal and idler photons. For all our experimental results, we see excellent agreement with the theoretical predictions. 

There are multiple areas for further improvements that would expand our SiP shaper's  utility in QIP.
Fundamentally, the complexity of the waveforms any Fourier-transform pulse shaper can produce---i.e., its time-bandwidth product---is determined by the number of independently controllable spectral elements~\cite{Weiner2000, Weiner2011}, which in our case is simply the number of filter channels (six). For a microring-based pulse shaper, the number of independent features is ultimately capped by the number of channels that can be packed within a single FSR ($\omega_\mathrm{FSR}/\Delta\omega$), which, at the 115~GHz FSR and 3~GHz spacing in our demonstration, implies that approximately 38 channels could be supported by simply adding more rings and phase shifters. 
Furthermore, although we have focused on phase-only shaping here, the addition of line-by-line amplitude control should be possible by incorporating variable optical attenuators inline with each phase shifter, components likewise readily available in silicon photonics~\cite{park2010influence,Ma2016}.

The performance of the shaper, including both spectral resolution and insertion loss, can be enhanced through careful design adjustments. For a single add-drop microring, the insertion loss at the drop port is set by the overcoupling ratio $Q_i/Q_L$, where $Q_i$ is the intrinsic quality factor and $Q_L$ the loaded quality factor. By designing a microring with stronger coupling or by reducing its propagation loss, the ratio increases and the insertion loss is reduced.
Additionally, while our current demonstration already achieves a resolution of 3 GHz, notably finer than typical bulk pulse shapers, further refinement is possible by increasing $Q_L$. With a fixed overcoupling ratio $Q_i/Q_L$ (and thus shaper loss), enhancing the spectral resolution is possible by minimizing the round-trip optical loss within the microrings and tuning the coupling appropriately.

Such higher quality factors can be obtained by further device engineering on the silicon photonics platform, or through other material platforms with ultralow waveguide propagation losses like silicon nitride \cite{ji2021methods,liu2021high} and thin-film lithium niobate \cite{zhang2017monolithic, zhu2021integrated}, which could facilitate linewidths (and hence channel spacings) approaching a few tens of~MHz and enabling biphoton wavepacket durations of tens of ns. With access to such long time apertures, a pulse shaper of this form could be used to coherently shape photons for interactions with MHz-scale optical cavities or atomic memories~\cite{Lvovsky2009, Lei2023}. To date, photonic wavepacket shaping on these timescales has been the the purview of inherently lossy temporal electro-optic \emph{intensity} modulation~\cite{Kolchin2008, Liu2014}, whereas ultrafine resolution pulse shaping could in principle realize such temporal control with unitary \emph{phase-only} spectral filters---pointing to exciting opportunities for our system in quantum transduction.

This integrated pulse shaper should find a number of applications in all-photonic QIP as well. In the context of the quantum frequency processor, such a fine-resolution shaper could enable arbitrary frequency-bin transformations with higher dimensions than possible with bulk components, by increasing in the number of frequency bins accessible in smaller total bandwidth and thereby relaxing demands on high-speed electro-optic phase modulation~\cite{lu2022high, Nussbaum2022}. Capabilities like those demonstrated in Fig.~\ref{fig4} could be useful in nonlocal spectral coding or measurement schemes for quantum communications~\cite{lukens2014orthogonal, lee2014entanglement, harris2008nonlocal, sensarn2009observation}. In scenarios where photons from an entangled pair are routed to different users, local phase operations by one user can be successfully ``decoded'' when the conjugate phase is applied by the other user; see, for example, Fig.~\ref{fig4} (last plot) where strong correlations at zero delay are recovered when conjugate phase vectors are applied. Due to the narrow total bandwidth of the BFCs used, such phase coding applications could be realized with direct time-resolved measurement, compared to previous demonstrations where nonlinear upconversion processes were required~\cite{lukens2014orthogonal}. 

\begin{acknowledgments}
The authors thank Matthew van Niekerk and Stefan Preble for providing materials for wirebonding, and Carsten Langrock and Martin Fejer for providing the PPLN chips. Portions of this work were presented at CLEO 2024 as paper number FTu4F.4. Funding was provided by the the National Science Foundation (2034019-ECCS, 1747426-DMR) and the U.S. Department of Energy (ERKJ353).
\end{acknowledgments}

\appendix

\begin{figure*}[t!]
\centerline{\includegraphics[width=0.8\textwidth]{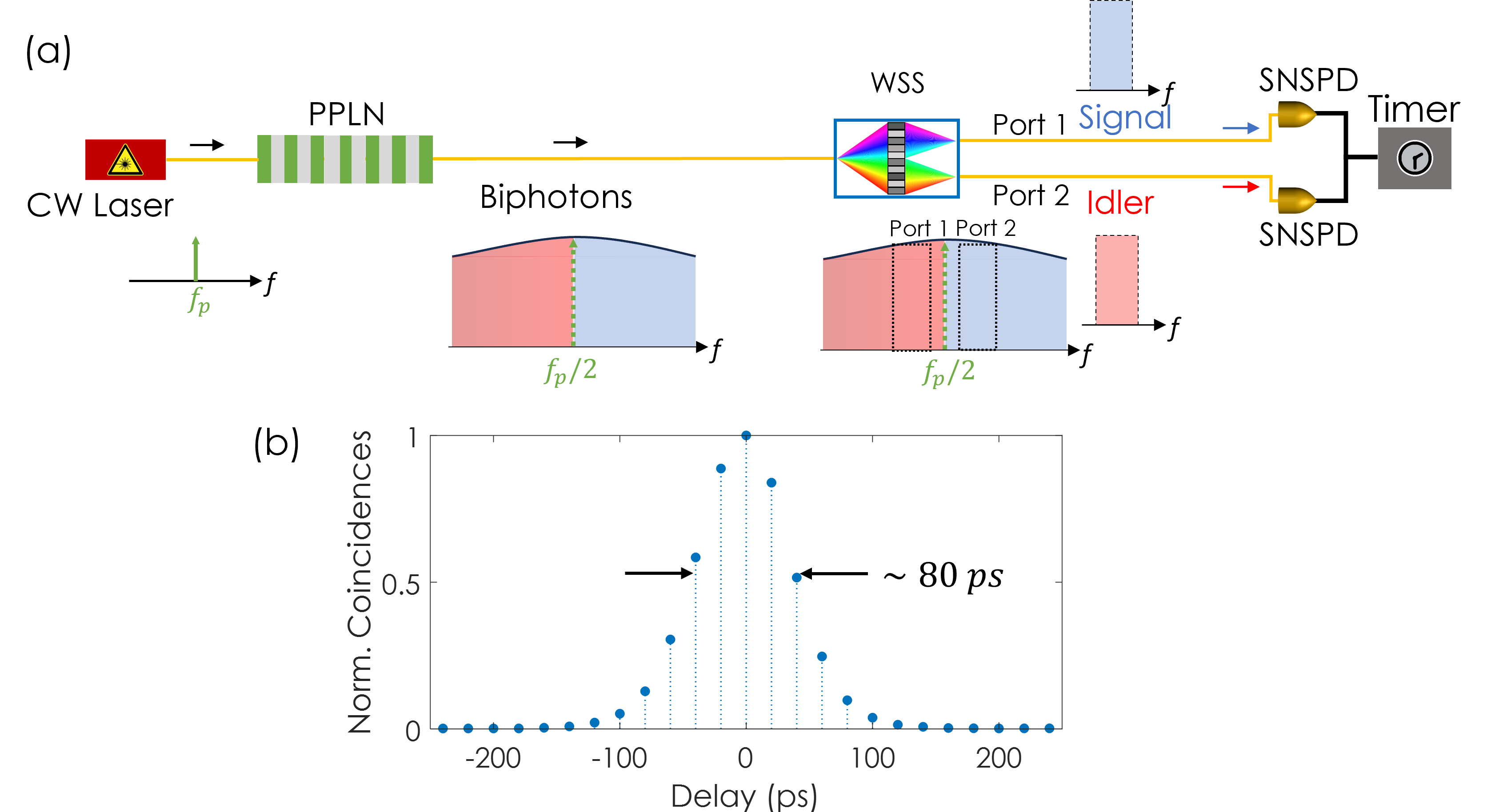}}
\caption{(a) Experimental setup for measuring the detection system impulse response.  CW: continuous-wave. PPLN: periodically poled lithium niobate. SNSPD: superconducting nanowire single-photon detector. WSS: wavelength-selective switch.
(b) Measured coincidence histogram for $50$~GHz-wide  WSS bins.}
\label{supp:fig1}
\end{figure*}

\section{Impulse Response Measurement}
\label{app:A}
As described in Eq.~(\ref{eq:convolve}), the impulse response of our detection system $h(\tau)$ is convolved with the ideal  correlation function $G^{(2)}(\tau)$ to obtain the measured coincidences $C(\tau)$. The impulse response is measured by the experimental setup depicted in Fig.~\ref{supp:fig1}(a). Biphotons are generated in the same manner as the main text, and the signal and idler spectra are directly carved and spatially separated by the WSS. The $50$~GHz-wide rectangular bins should produce a biphoton correlation function with an FWHM of $\sim$20~ps, well below our detector system jitter; thus the spread in coincidences will reflect the system response rather than the optical bandwidth. The measured correlation function is shown in Fig.~\ref{supp:fig1}(b), which possesses a Gaussian shape with an $\sim$80~ps FWHM that accounts for both detector and timing electronics jitter. This impulse response is used to produce the simulated traces in the main text.

\section{Spectral Phase Calibration}
\label{app:B}
\begin{figure*}[t]
\centerline{\includegraphics[width=\textwidth]{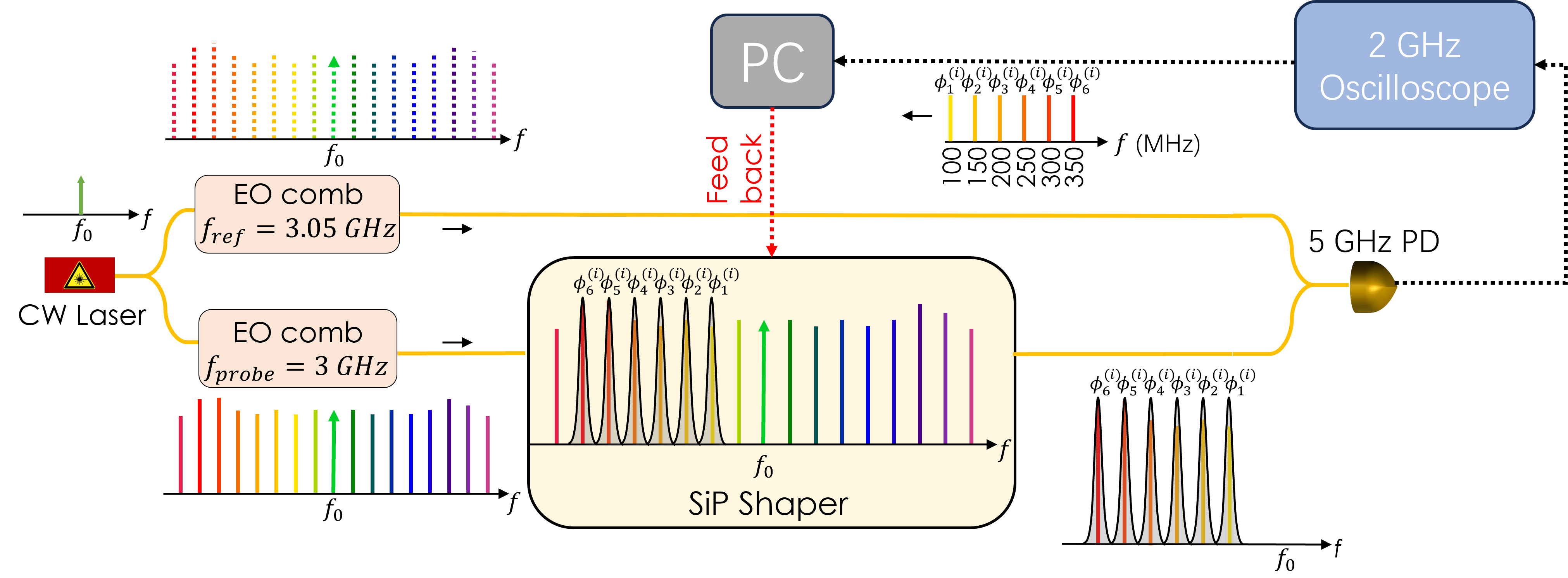}}
\caption{Schematic for the phase measurement performed in the experiments of Fig.~\ref{fig2}. $d = 6$ channels are shown here for convenience. CW: continuous-wave. PD: photodetector. PC: personal computer.}
\label{supp:fig2}
\end{figure*}

To calibrate the phases applied by the SiP shaper, we perform dual-comb spectroscopy using two electro-optic (EO) frequency combs, both pumped by a CW laser operating at 1551.8~nm (193.2~THz). The detailed procedures are described in \cite{Cohen2024nc}. In our setup, each EO comb generator consists of an intensity modulator followed by a phase modulator. One EO comb (``probe'') is driven at a repetition rate  equal to the channel spacing, $f_\mathrm{probe} = 3$~GHz, while the other EO comb (``reference'') is driven at a slightly higher repetition rate of $f_\mathrm{ref} = 3.05$~GHz, creating an offset of $50$~MHz. When these two EO combs are combined and measured by a photodetector (PD), they generate beat frequencies at multiples of the 50~MHz offset. These beat notes are captured by a real-time oscilloscope (Rohde \& Schwarz RTO1024). By performing a fast Fourier transform (FFT) on the time samples, we can extract the amplitude and phase of the beat notes, allowing us to infer the phase difference between the corresponding EO comb lines. 

Prior to dual-comb spectroscopy, we characterize the initial spectral phase of both EO combs~\cite{Cohen2024nc}. Since the probe and reference combs are driven with similar RF power and experience negligible dispersion, with the only difference being the 50~MHz offset in repetition rate, their spectral phases are nearly identical. Consequently, the phase information obtained from the FFT of the beat notes can be directly attributed to the phases applied by the shaper chip to the probe comb lines.

\subsection{Shared Signal-Idler Filters} \label{subsec:1}
Figure~\ref{supp:fig2} depicts the phase measurement setup for the experiment of shared signal-idler filters in Sec.~\ref{sec:shared}. 
We tune the CW laser wavelength and send the probe comb to the SiP shaper such that the $2^\mathrm{nd}-7^\mathrm{th}$ comb lines on the low-frequency side of the pump pass through the six programmed channels. This setup results in six beat notes in the FFT, starting at $100$~MHz spaced by $\Delta f_\mathrm{rep} = 50$~MHz. Finally, based on the phase measurement, the drive signals to the shaper phase shifters are adjusted to achieve the desired phase vector. Because any global linear phase is equivalent to a common delay and hence unobservable in the biphoton correlation function, the zero-phase condition [$\phi_k^{(s)}=\phi_k^{(i)}=0$] can be defined with two of the phase shifters fixed arbitrarily and any additional phase shifters tuned to achieve a linear slope in the FFT spectrum.

\begin{figure*}[t!]
\centerline{\includegraphics[width=\textwidth]{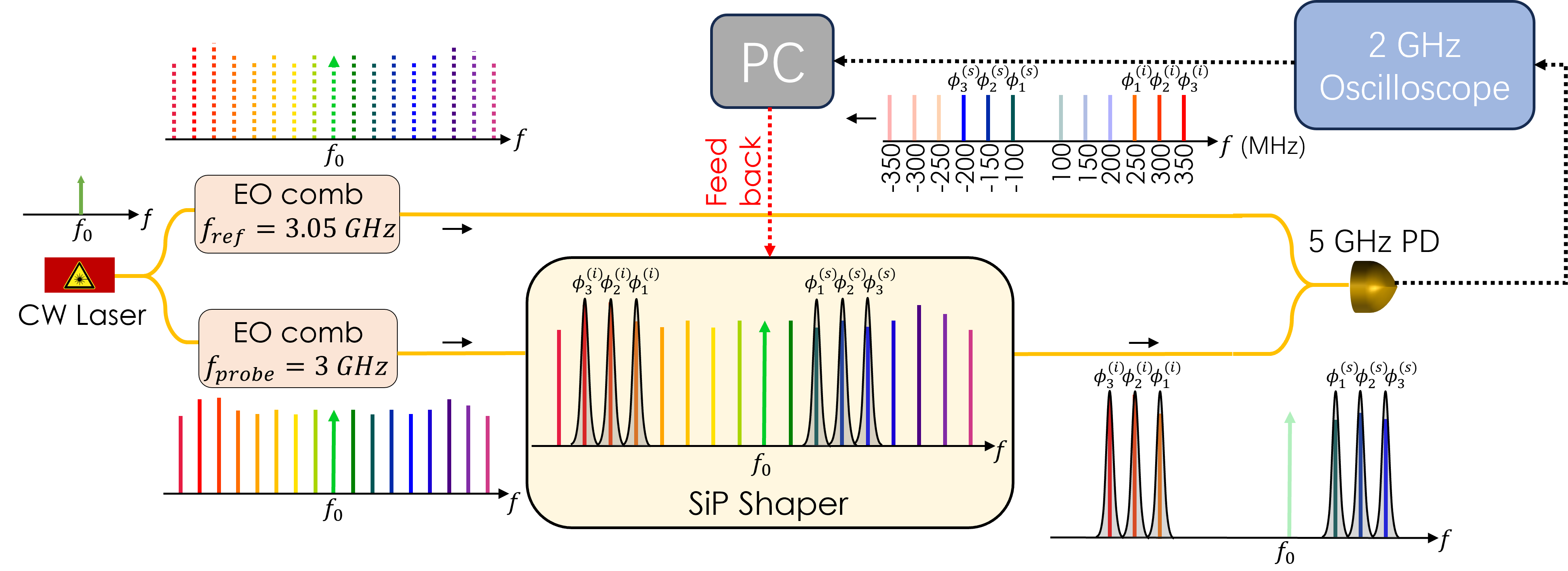}}
\caption{Schematic for the phase measurement performed in the experiments of Fig.~\ref{fig3} and Fig.~\ref{fig4}. PD: photodetector. PC: personal computer.}
\label{supp:fig3}
\end{figure*}

\subsection{Distinct Signal-Idler Filters}

\begin{figure*}[h!]
\centerline{\includegraphics[width=\textwidth]{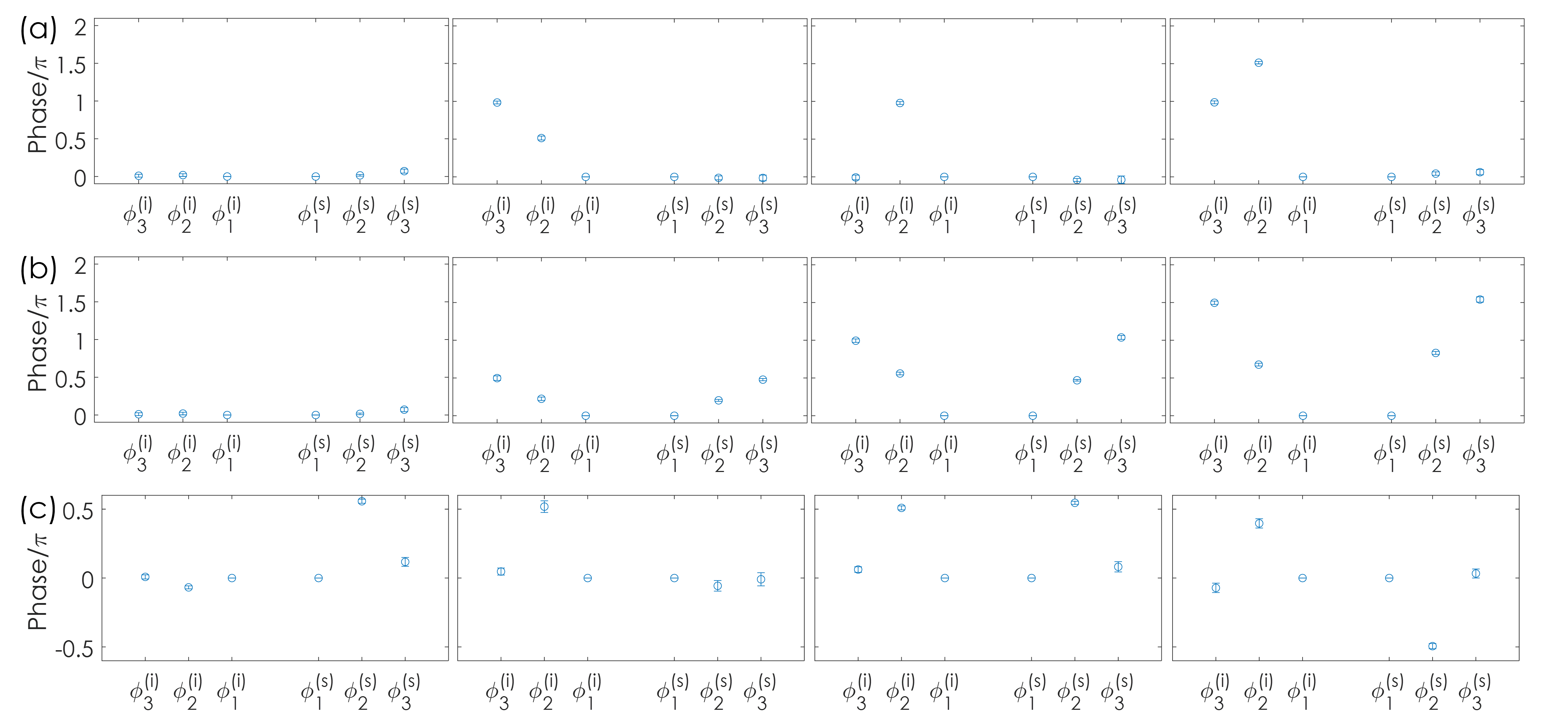}}
\caption{Measured phases applied across bins corresponding to distinct signal-idler filter experiment in (a) Fig.~\ref{fig3}(c), (b)~Fig.~\ref{fig3}(d), and (c)~Fig.~\ref{fig4}.}
\label{supp:fig4}
\end{figure*}

The methods are modified slightly for the remaining results in Sec.~\ref{ExpResults} where we manipulate the correlation with independent control of signal and idler phases. In this case we program two $d=3$ subcombs with a $3$~GHz bin spacing and $27$~GHz center-to-center subcomb spacing. We align the CW laser wavelength such that the probe comb lines passing through the chip are as shown in Fig.~\ref{supp:fig3}: the $2^\mathrm{nd}$--$4^\mathrm{th}$ comb lines on the high-frequency side and the $5^\mathrm{th}$--$7^\mathrm{th}$ on the low-frequency side. Because the two subcombs are on opposite sides of the common pump, we extract the phases at beats $-$200, $-$150, $-$100, 250, 300, and 350 ~MHz for $\phi_3^{(s)}$, $\phi_2^{(s)}$, $\phi_1^{(s)}$, $\phi_1^{(i)}$, $\phi_2^{(i)}$, and $\phi_3^{(i)}$, respectively (see Fig.~\ref{supp:fig3}).  
As before, any global linear phase is unobservable in the biphoton histogram, so two of the six phase shifts can be set arbitrarily; specifically, we choose $\phi_1^{(i)}$ and $\phi_1^{(s)}$ as reference phases and subtract off the slope they define for all phase shifts reported in Fig.~\ref{supp:fig4}, which shows all the applied phases (measured after programming them) adopted for distinct signal-idler filter experiments in Sec. \ref{sec:distinct} and \ref{sec:nonlinear}.

\clearpage

%

\end{document}